\documentclass[twocolumn,prb,floats,showpacs,superscriptaddress]{revtex4}
\usepackage{graphicx,epsfig}
\usepackage{times}
\usepackage{graphics,dcolumn,bm,float}
\usepackage{amssymb,amsmath,rotate,color}

\begin{document}
\unitlength 1 cm
\newcommand{\be}{\begin{equation}}
\newcommand{\ee}{\end{equation}}
\newcommand{\bearr}{\begin{eqnarray}}
\newcommand{\eearr}{\end{eqnarray}}
\newcommand{\nn}{\nonumber}
\newcommand{\vk}{\vec k}
\newcommand{\vp}{\vec p}
\newcommand{\vq}{\vec q}
\newcommand{\vkp}{\vec {k'}}
\newcommand{\vpp}{\vec {p'}}
\newcommand{\vqp}{\vec {q'}}
\newcommand{\bk}{{\bf k}}
\newcommand{\bp}{{\bf p}}
\newcommand{\bq}{{\bf q}}
\newcommand{\br}{{\bf r}}
\newcommand{\bR}{{\bf R}}
\newcommand{\up}{\uparrow}
\newcommand{\down}{\downarrow}
\newcommand{\fns}{\footnotesize}
\newcommand{\ns}{\normalsize}
\newcommand{\cdag}{c^{\dagger}}
\newcommand{\degree}{\ensuremath{^\circ}}

\title{Charge-doping-induced variation of BaFe$_2$As$_2$ electronic structure and the emerging physical effects }
\author{Mehdi Hesani}\email{mehdi.hesani@gmail.com}
\author{Ahmad Yazdani}
\author{Kourosh Rahimi}
\affiliation{Condensed Matter Group, Department of Basic Sciences, Tarbiat Modares University, Jalal-Ale-Ahmad Avenue, Tehran, Iran }

\date{\today}

\begin{abstract}
We studied the relationship between the charge doping and the correlation, and its effects on the spectral function of the BaFe$_2$As$_2$  compound in the framework of the density functional theory combined with the dynamical mean field theory (DFT+DMFT). The calculated mass enhancements showed that the electronic correlation varies systematically from weak to strong when moving from the heavily electron-doped regime to the heavily hole-doped one. Since the compound has a multi-orbital nature, the correlation is orbital-dependent and it increases as hole-doping increases. The Fe-3d$_{xy}$ (xy) orbital is much more correlated than the other orbitals, because it reaches its half-filled situation and has a narrower energy scale around the Fermi energy. Our findings can be consistently understood as the tendency of the heavily hole-doped BaFe$_2$As$_2$   compound to an orbital-selective Mott phase (OSMP). Moreover, the fact that the superconducting state of the heavily hole-doped BaFe$_2$As$_2$   is an extreme case of such a selective Mottness constrains the non-trivial role of the electronic correlation in iron-pnictide superconductors. In addition, the calculated spectral function shows a behavior that is compatible with experimental results reported for every charge-doped BaFe$_2$As$_2$   compound and clarifies the importance of the characterization of its physical effects on the material.  
\end{abstract}

\maketitle
\section{Introduction}
 
Since the discovery of superconductivity in the iron-based compound of LaO$_{1-x}$F$_x$FeAs with the transition temperature (T$_c$) of 26 K [1], a wide variety of crystal structures has been reported with this property  [2–5]. One of the structures is AFe$_2$As$_2$ [A=Ba, Co, Sr], known as the “122” group. In particular, the ground state of  BaFe$_2$As$_2$ (as the parent compound) is antiferromagnetic (AF). However, doping electrons (as BaFe$_{2-x}$Co$_x$As$_2$ [2]), doping holes (as Ba$_{1-x}$K$_x$Fe$_2$As$_2$ [3]), or applying pressure [6] suppress the AF state and develops a superconducting state. Hole-doped Ba$_{1-x}$K$_x$Fe$_2$As$_2$ compounds are attractive for further investigations, because they have exhibited various phenomenon, i.e. a weak but distinct heat capacity jump at T$_c$ when  $x\approx0.7$  [7] and an abrupt change in the energy gap when  $x\geq0.6$  [8]. The BaFe$_2$As$_2$ compound has been studied by angle-resolved spectroscopy (ARPES), for almost all electron- and hole-doping amounts [8\text{-}20]. The experimental electronic structures of the material are somewhat inconsistent in terms of some details, which is probably due to the incoherent states of the material  [20] caused by its electronically correlated behavior. There are several theoretical studies on the compounds, but they are related to a specific region and often inconsistent with experimental results. It is very important to study the electronic structures of the materials systematically for the full range of charge doping.

From weakly- to strongly-correlated conduction electrons have been observed in iron-based superconductors. Doping electrons or holes strongly affects the correlation in the materials. Theoretical and experimental investigations of the materials have shown that weakly- and strongly-correlated electrons coexist in much of the phase diagrams, and the differentiation between them gets stronger by hole-doping [20\text{-}24]. The Hund$'$s coupling effects of decoupling of the charge excitation  in different orbitals cause the selective orbital behavior. The behavior of every orbital depends on how much it is doped above the half-filled situation. Approaching the half-filled situation mainly increases the correlation in the orbital. The orbital-selective Mott phase (OSMP) would be realized for half-filled conduction bands [21]. In this scenario, each orbital behaves as a Mott insulator, with a correlation strength that depends on its character. This selective Mottness could explain the physics of iron-based superconductors and their similarity to cuprate ones.

The electronic correlation strength and its role in iron-based superconductors are still under debate. A wide range of correlations, from weak [25-27] to strong  [28-29], has been investigated for these materials in theoretical studies. The moderate correlation region, where electrons are neither fully itinerant nor fully localized, can best describe the multi-orbital nature of the materials. An intriguing theory proposed by Haule and Kotliar  [30], suggesting that the Hund$'$s rule coupling is indeed responsible for the correlation, has a good agreement with the experimental results [22,31]. The method combining the density functional theory with the dynamical mean field theory (DFT+DMFT) is very successful in predicting the magnetic, structural, superconductivity, and electronic properties of these systems [22,31\text{-}33]. The combined method emphasizes the crucial role of the correlation in these materials.

We focused here on the interrelationship between the charge-doping and the correlation in the  BaFe$_2$As$_2$ compound in the framework of the density functional theory combined with the dynamical mean field theory (DFT+DMFT). The spectral function and the occupation number (ON) of Fe-3d orbitals for various charge-doping situations were extracted from the performed calculations, which indicated the important role of the Fe-3d$_{xy}$ (xy) orbital in superconductivity of these materials through the doping process.

\begin{figure*}
\centerline{\includegraphics[width=1.0\linewidth]{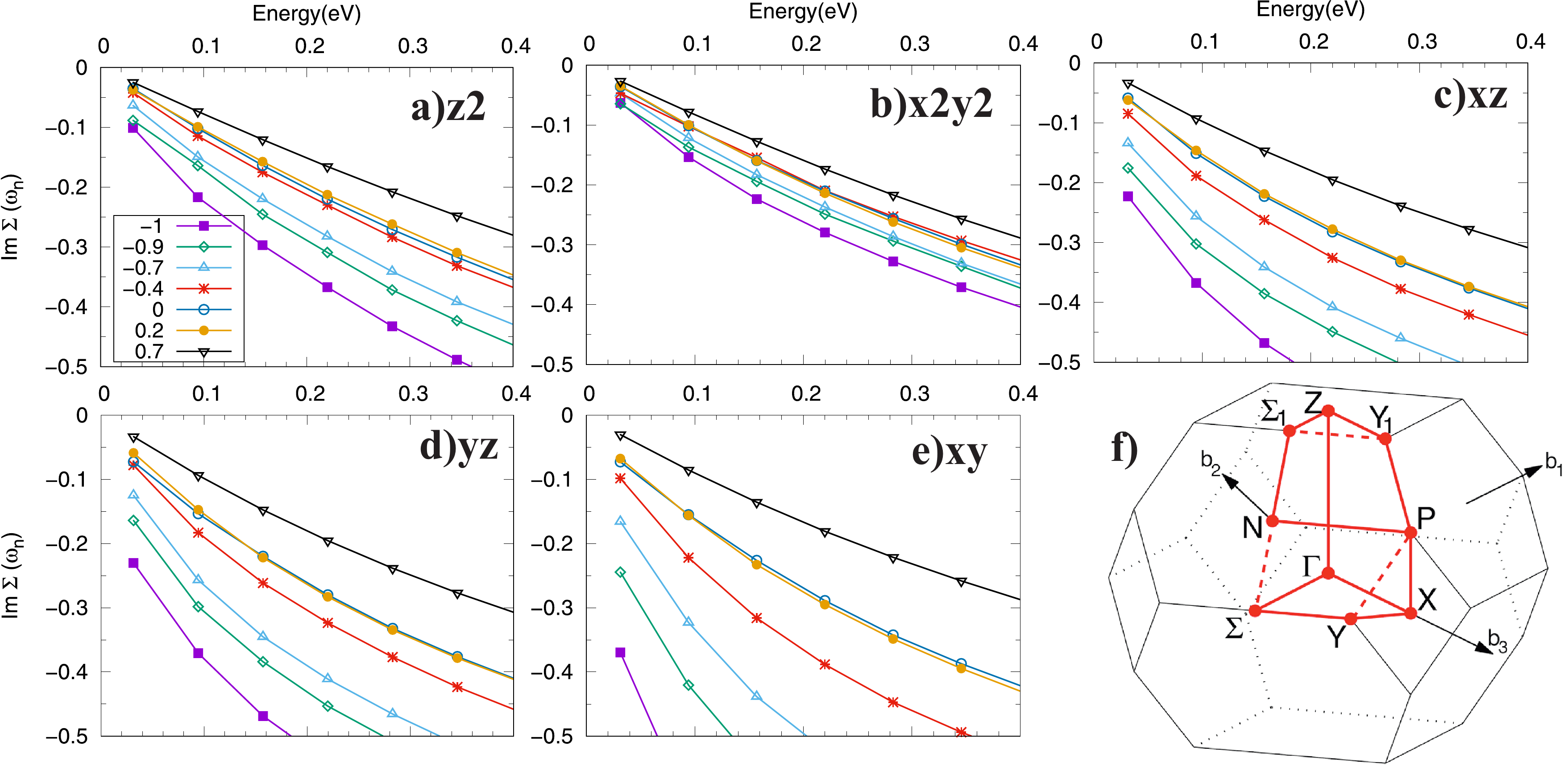}}
\caption{(Color online) The charge-doping dependent imaginary part of the self-energy $\operatorname{Im}\Sigma (i{{\omega }_{n}})$) of Fe-3d orbitals (a) z2, (b) x2−y2, (c) xz, (d) yz, and (e) xy at low energies for Ba($x$), where $x$ denotes the doping amount (in terms of e/cell). (f) The Brillouin zone  with high-symmetry labels of the body-centered tetragonal lattice. }
\end{figure*}

\section{COMPUTATIONAL DETAILS}

We used the charge self-consistent combination of the density functional theory with the dynamical mean field theory (DFT+DMFT) [34], implemented in two full-potential methods: the augmented plane-wave method and the linear muffin-tin orbital method (as in the WIEN2K package  [35]). The DFT calculations were performed in the WIEN2K package [27] in its generalized gradient approximation [Perdew-Burke-Ernzerhof (PBE)-GGA]  [36]. The self-energy was added to the DFT results, and DMFT was applied to them. The computations were converged with respect to the charge density, the total energy, and the self-energy. The continuous-time quantum Monte Carlo (CTQMC) method  [37,38] in its fully rotationally-invariant form was used to solve the quantum impurity problem. All 3d orbitals, i.e. d$_{3z2-r2}$ (z2), d$_{x2-y2}$ (x2y2), d$_{xz}$ (xz), d$_{yz}$ (yz), and d$_{xy}$ (xy), were considered as correlated for the Fe site. The self-energy was analytically extended from the imaginary axis to the real axis using the maximum entropy method.

A virtual crystal approximation were used for various electron(e)-doped BaFe$_{2-x}$Co$_x$As$_2$ (Ba($x$), $x$ varies from 0 to 2 e/cell) or hole(h)-doped  Ba$_{1-x}$K$_x$Fe$_2$As$_2$ (Ba($x$), $x$  varies from 0 to -1 e/cell). The experimentally determined lattice structures  [39-42] were used for the calculations. All structural positions were relaxed by the DFT+DMFT method  to 0.01 eV/${\AA}$, as reported by Haule and Pascut  [33] for paramagnetic states. The same Coulomb interaction of $U =5.0$ eV and the same Hund’s coupling of $J_H =0.80$ eV were used in all of our DFT+DMFT calculations  [22]. The fine 16$\times$16$\times$7 k-point mesh and totally 3$\times$10$^9$ Monte Carlo steps for each iteration were used for the paramagnetic phase of the  BaFe$_2$As$_2$ compound for various charge-doping situations at 120 K.

\begin{table}
\caption{The occupation numbers of the Fe-3d orbitals for various charge-doping situations.}
\begin{tabular}{l*{6}{c}}
\hline
e${\diagdown}$Character              & 3d &  z2 &x2y2&xz&yz&xy \\
\hline
-1 &6.04 & 1.31 & 1.25 & 1.16 & 1.16 & 1.14  \\
-0.9&6.08 & 1.35 & 1.21 & 1.17 & 1.17 & 1.17  \\
-0.8&6.12 & 1.33& 1.20 & 1.18 & 1.18 & 1.22 \\
-0.7 &6.12 & 1.33 & 1.22 & 1.18 & 1.18 & 1.20  \\
-0.6 &6.13 & 1.33 & 1.20 & 1.19 & 1.19 & 1.23  \\
-0.5 &6.12 & 1.31 & 1.21 & 1.19 & 1.19 & 1.22  \\
-0.4 &6.15 & 1.32 & 1.21 & 1.19 & 1.19 & 1.23  \\
-0.25 &6.19 & 1.32 & 1.21 & 1.21 & 1.21 & 1.25  \\
-0.17 &6.21 & 1.32 & 1.20 & 1.21 & 1.21 & 1.26  \\
-0.1 &6.23 & 1.32 & 1.21 & 1.21 & 1.21 & 1.26  \\
0 &6.28 & 1.32 & 1.20 & 1.24 & 1.23 & 1.29  \\
0.14 &636 & 1.33 & 1.22 & 1.25 & 1.25 & 1.30  \\
0.22 &6.38 & 1.34 & 1.22 & 1.25 & 1.25 & 1.31  \\
0.4 &6.49 & 1.35 & 1.22 & 1.28 & 1.28& 1.36 \\
0.6&654& 1.35 & 1.20 & 1.30 & 1.30 & 1.39  \\
0.7 &6.57 & 1.35 & 1.19 & 1.31 & 1.31 & 1.40  \\
\hline
\end{tabular}

\end{table}

\section{RESULTS AND DISCUSSION }

\begin{figure}
\centerline{\includegraphics[width=1.0\linewidth]{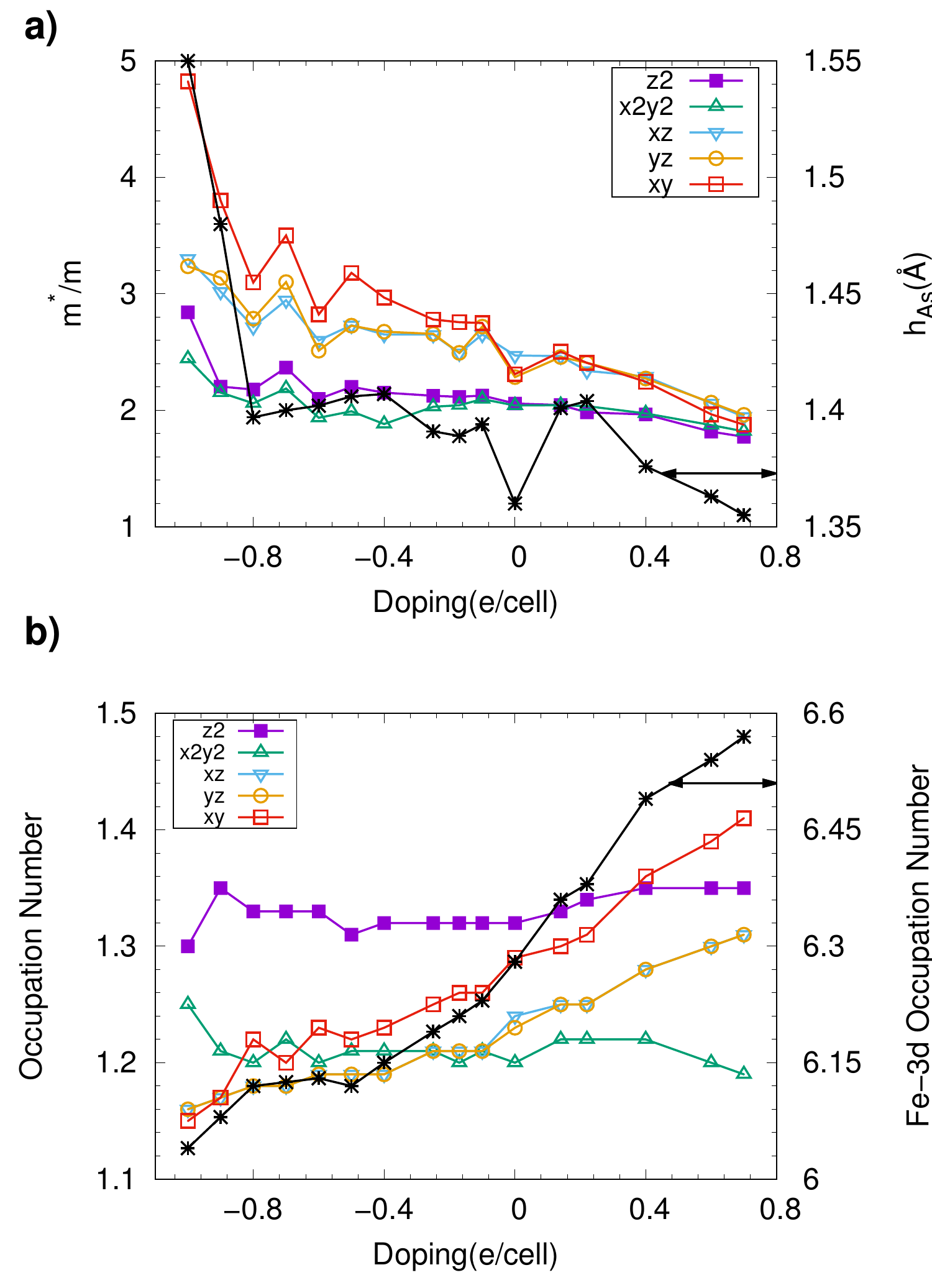}}
\caption{(Color online) (a) The mass enhancement of different Fe-3d orbitals (left) and the anionic height (h$_{As}$) (right) show a similar trend across various e-doping situations. (b) The occupation number of the Fe-3d orbitals for various e-doping situations (left). The Fe-3d occupation number (right) (the black vector points towards the corresponding axis).}
\end{figure}

The charge-doping dependent imaginary part of the self-energies ($\operatorname{Im}\Sigma (i{{\omega }_{n}})$) ) of Fe-3d orbitals for Ba($x$) is shown in ‎Figure 1, indicating that $\operatorname{Im}\Sigma$ is very orbital-selective, and the orbital-selectivity is enhanced by increasing the extra holes in the system. As the e-doping increases, the $\operatorname{Im}\Sigma$ behavior becomes more linear at low energies, with a slope directly related to the quasiparticle mass enhancement, as expected from the Fermi liquid theory. Increasing the e-doping enhances the coherence scale in these materials. In the Fermi liquid theory, the inverse quasiparticle (QP) lifetime is defined as the scattering rate $\Gamma =-Z\operatorname{Im}\Sigma (i\omega )$ , where $Z={{({{m}^{*}}/m)}^{-1}}={{(1-\frac{\partial \Sigma (i\omega )}{i\omega })}^{-1}}{{|}_{i\omega \to i{{0}^{+}}}}$  can be interpreted as the quasiparticle weight and $\operatorname{Im}\Sigma (i{{0}^{+}})$   is the imaginary part of the self-energy at the zero frequency. At the zero temperature, where $\operatorname{Im}\Sigma (i{{0}^{+}})\to 0$   , the system is in the coherent phase with an infinite QP lifetime. At a constant temperature, $\operatorname{Im}\Sigma (i{{0}^{+}})$ and consequently the QP lifetime are both finite. Therefore, in order to identify the coherent QPs, the temperature has to be lower than the QP width, i.e. $T<-Z\operatorname{Im}\Sigma (i{{0}^{+}})$  . At $x=-0.4$, for z2, x2y2, xz, yz, and xy, we have $\operatorname{Im}\Sigma (i{{0}^{+}}) =42, 46, 85, 78,$ and $98$ meV as well as $Z=0.46, 0.53, 0.38, 0.37,$ and $0.34$, and hence $-Z\operatorname{Im}\Sigma (i{{0}^{+}})  =227, 288, 370, 337$, and $384$ K, respectively. We performed all calculations at 120 K, which is lower than the QP width corresponding to all Fe-3d orbitals, which means that no coherent quasi-particle can be expected for them. Our results are consistent with the work by Werner et al. [20], where they found that the BaFe$_2$As$_2$ compound is in its incoherent state at its optimum h-doping (i.e. 0.4 h/cell ). Our results show that for h-doping amounts more than the optimum amount, the materials would be at their incoherent states. This suggests that a probable coherent-incoherent crossover would happen in BaFe$_2$As$_2$  through the h-doping. Hardy et al.  [23] reported a robust experimental evidence for the coherence-incoherence crossover in KFe$_2$As$_2$ , which is confirmed by our calculations. It is a strong evidence of the strongly correlated behavior of the material, similar to what is found in heavy fermions due to the Hund$'$s coupling between orbitals.

The variation of the orbital-dependent effective masses for Ba($x$), as shown in ‎Figure 2 (a), follows the variation of the anionic height (the distance between As and the Fe-plane) for various charge-doping situations, which confirms the sensitivity of the correlation existing in these materials to the anionic height (h$_{As}$). An increase in the anionic height more than 1.40 ${\AA}$ affects the correlation severely and enhances the effective mass  (m$^*$/m) to 5 for the xy orbital. The highest variation in the mass enhancement is for the xy orbital, which varies from 1.4 to 5 as $x$ varies from 0.7 to -1. The finding is consistent with the imaginary self-energy, which shows that the materials exhibit an orbital-selective behavior. The differentiation between the orbitals becomes stronger by increasing the holes in the system, meaning that the xy orbitals show a very strong correlation, while the others behave less correlated.

To investigate the variation of the Fe-3d orbitals, we computed the occupation number (ON) of Fe-3d orbitals for various charge-doping situations. ‎Table 1 and ‎Figure 2 (b) present the variation of ONs of these orbitals for various charge-doping situations. In our calculations regarding the h-doping, the variations of the ONs of Fe-3d orbitals and the xy orbital are 0.24 and 0.14, respectively, indicating that nearly 60 percent of the variation is for the xy orbital. The finding emphasizes the important role played by the xy orbital in these materials. Losing the carrier concentration in the orbital could bring it to the half-filled situation, which explains the enhancement of the correlation in the xy orbital. Approaching the half-filled situation increases the possibility of OSMP for the heavily h-doped BaFe$_2$As$_2$. In this situation, it seems that the xy orbital independently behaves as a Mott insulator, while the other orbitals are at their metallic phases. Our results suggest the tendency of iron-pnictide superconductors to OSMP. The tendency to OSMP has been argued mostly in chalcogenides  [24,43-45]. However, its occurrence has also been recently discussed in pnictides [21,23,46].

\begin{figure*}
\centerline{\includegraphics[width=1.0\linewidth]{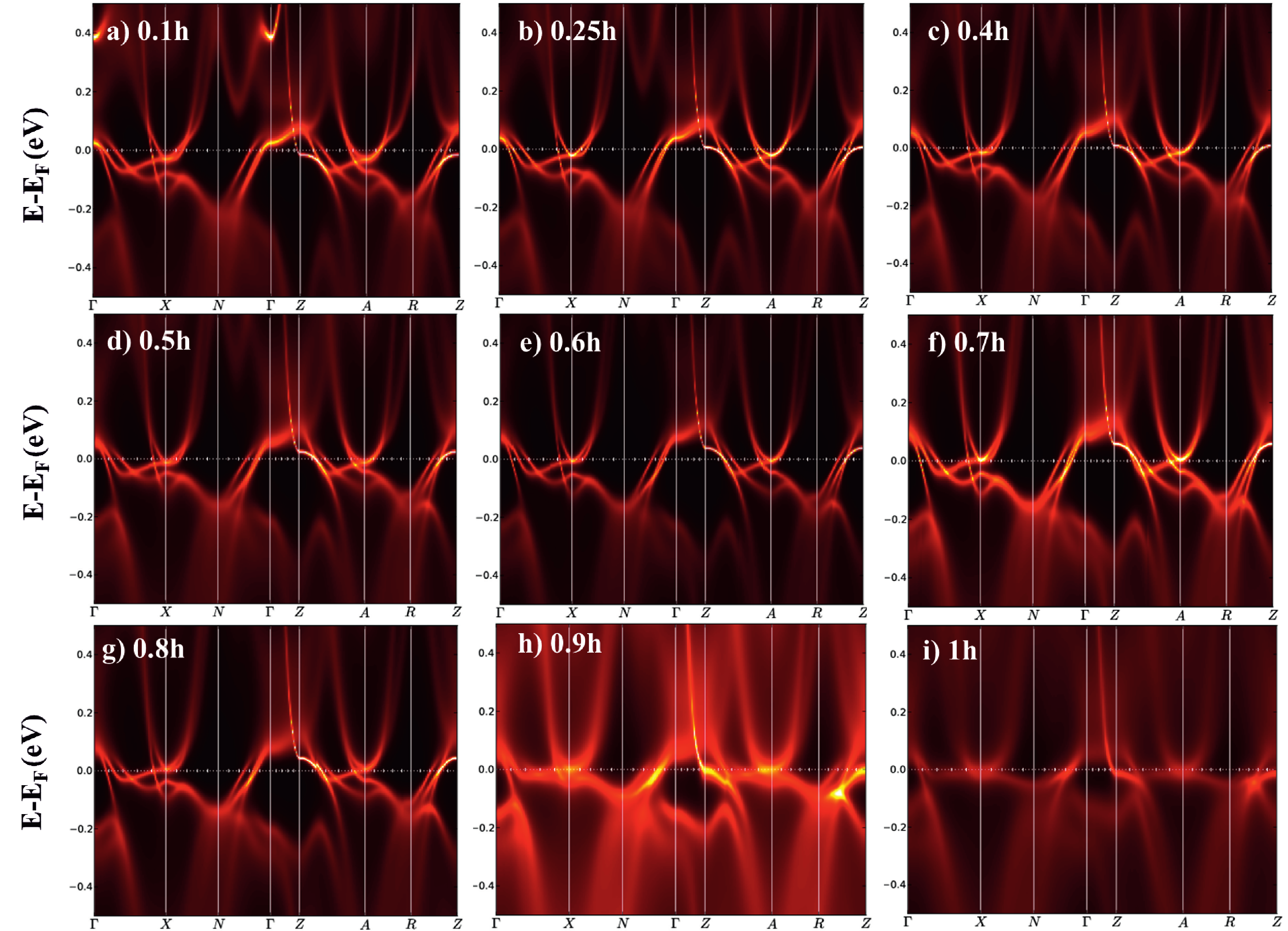}}
\caption{(Color online) The spectral functions of BaFe$_2$As$_2$ computed by the DFT+DMFT method for various h-doping situations: (a) 0.1, (b) 0.25, (c) 0.4, (d) 0.5, (e) 0.6, and (f) 0.7, (g) 0.8, (h) 0.9, and (i) 1.0 h/cell. The Brillouin zone with high-symmetry points is shown in Figure 1(f).}
\end{figure*}

\begin{figure*}
\centerline{\includegraphics[width=0.8\linewidth]{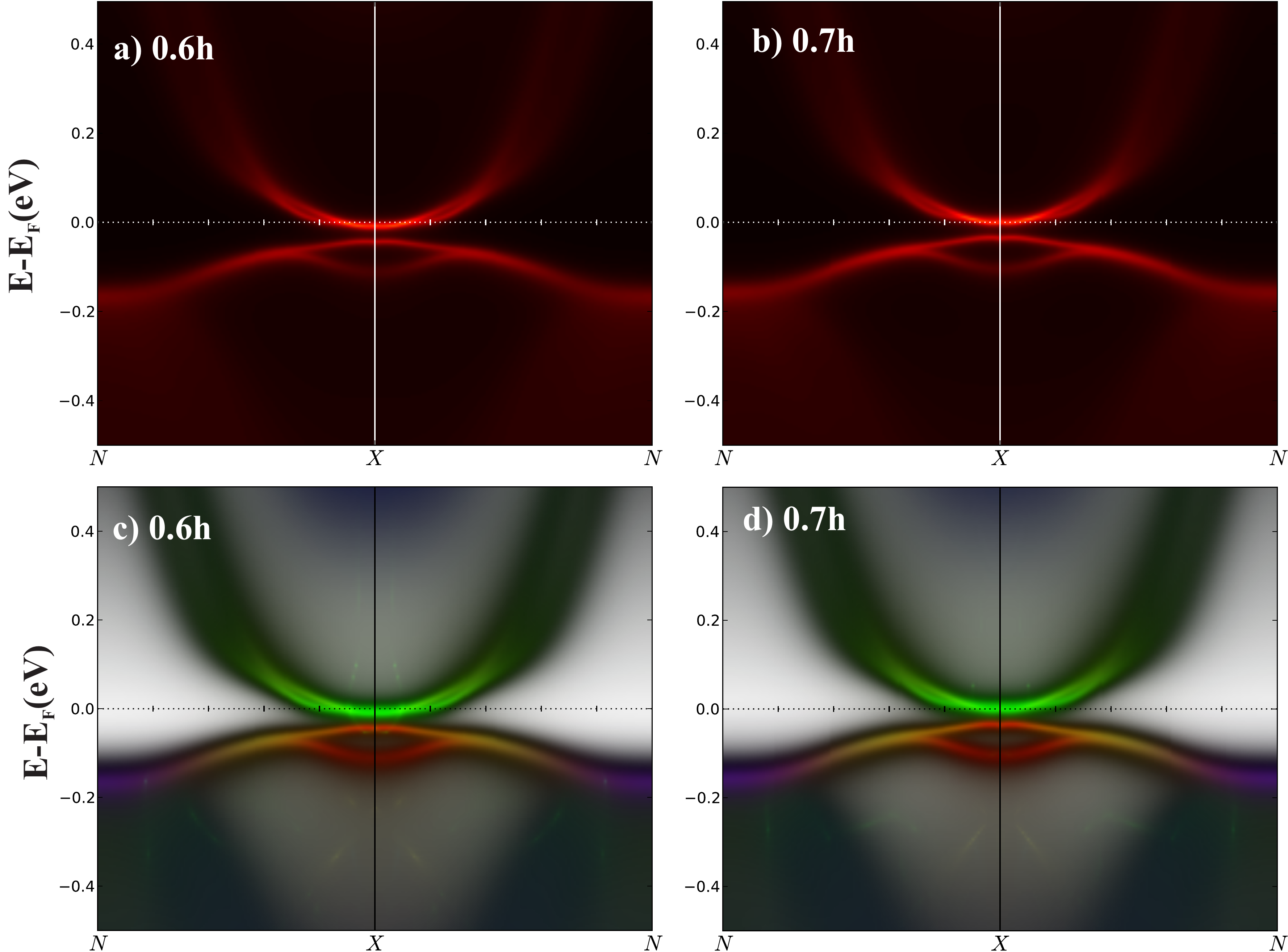}}
\caption{(Color online). The spectral functions of BaFe$_2$As$_2$ computed by DFT+DMFT along the  $N-X-N$ path (Figure 1(f)) for (a) 0.6 and (b) 0.7 h/cell, and their corresponding orbital-resolved spectral functions computed by DFT+DMFT for (d) 0.6 and (e) 0.7 h/cell (z2 and x2−y2 are in blue, xz and yz are in green, and xy is in red).}
\end{figure*}

In the e-doped region, the mass enhancement shows a significant decrease through the collapsing of the anionic height by increasing the e-doping in the system. We could classify the heavily e-doped region as weakly correlated. The collapse of the anionic height plays an important role here. The variation of ON in this region is faster than in the former region (the h-doping region). The variations of ON for xz, yz, and xy orbitals are significant, while they are negligible for z2 and x2y2 orbitals. The former variations have the main contribution to the Fermi surface, which makes it clear that the hole pockets at the Fermi surface would shrink or disappear by increasing the e-doping in the system. We discussed this issue in the following.

The spectral functions of BaFe$_2$As$_2$ computed by the DFT+DMFT method for various h-doping situations are plotted in ‎Figure 3. The spectral functions corresponding to low carrier-doping show three hole pockets at the center of the Brillouin zone and two electron pockets at the zone corner, which is confirmed by previous DFT [47] and DFT+DMFT [22] research. It is also compatible with ARPES measurements [8,48]. At the center, the outer band exhibits mostly the xy character. The middle band exhibits a combination of the xz/yz and x2y2 characters, and the inner band exhibits mostly a combined xz/yz character, as it is illustrated in ‎Figure 5 (d). Increasing the hole-doping affects mostly the outer band and makes it wider, meaning that the outer hole pocket at the Fermi surface becomes very wide. On the other hand, this band becomes narrower around the Fermi energy (E$_F$), which could explain the enhancement of the correlation in the xy orbital by increasing the hole-doping, especially for 0.9 and 1 h/cell situations (‎Figure 3). The xy mass renormalization is much stronger than that of other orbitals, which is due to its narrower energy scale. The results are consistent with our previous results on the mass enhancement and the occupation number. The middle and the inner hole pockets become wider for more hole-doping amounts. Nevertheless, they are not as narrower as the outer pocket. Therefore, the xz/yz orbitals through losing their carrier are not as correlated as the xy orbital.

\begin{figure*}
\rightline{\includegraphics[width=1.0\linewidth]{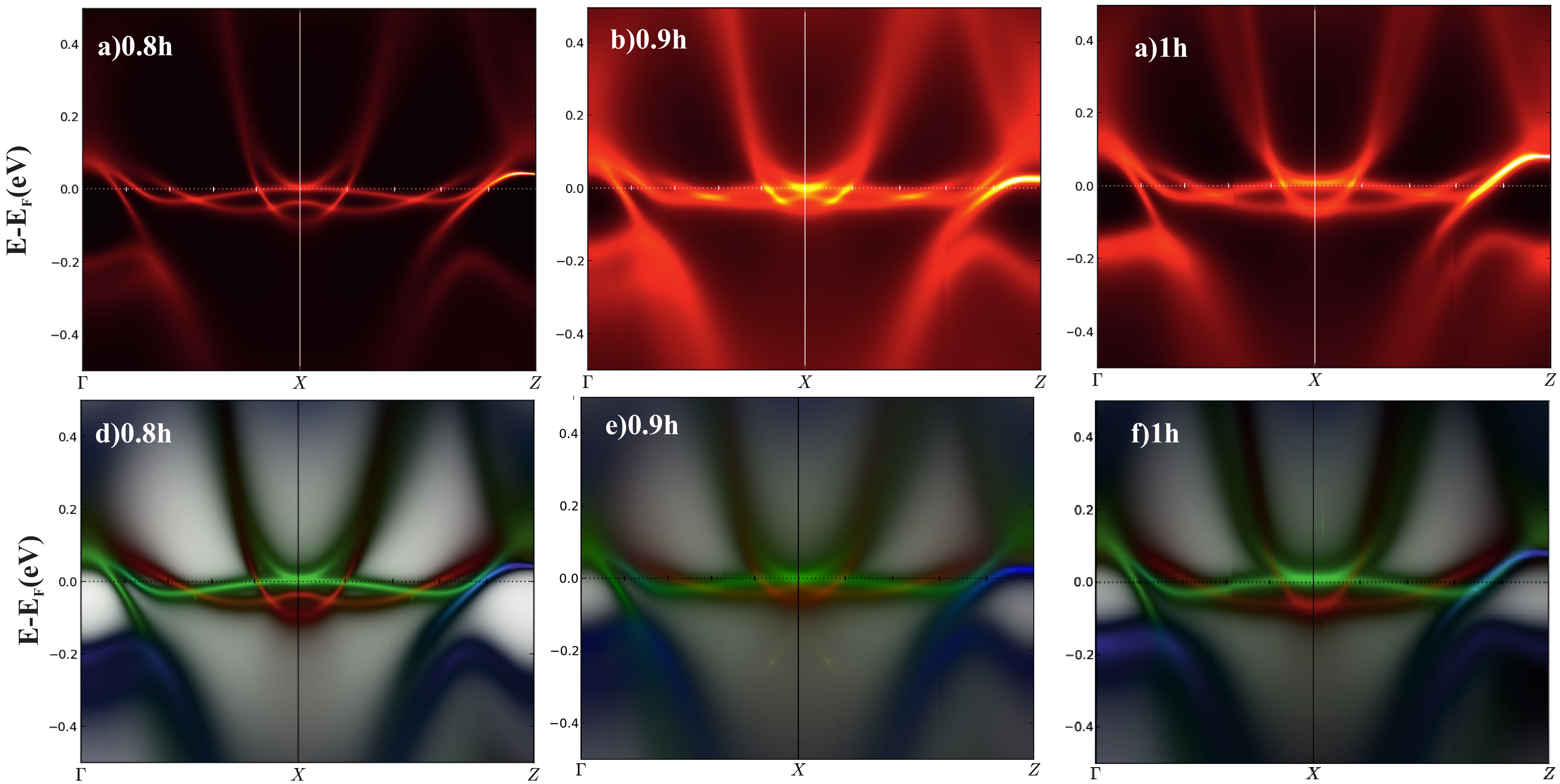}}
\caption{(Color online). The spectral functions of BaFe$_2$As$_2$ computed by DFT+DMFT along the $\Gamma-X-Z$ path (Figure 1(f)) for (a) 0.8, (b) 0.9 and (c) 1 h/cell, and their corresponding orbital-resolved spectral functions computed by DFT+DMFT for (d) 0.8, (e) 0.9, and (f) 1 h/cell (z2 and x2−y2 are in blue, xz and yz are in green, and xy is in red).}
\end{figure*}

\begin{figure}
\centerline{\includegraphics[width=0.9\linewidth]{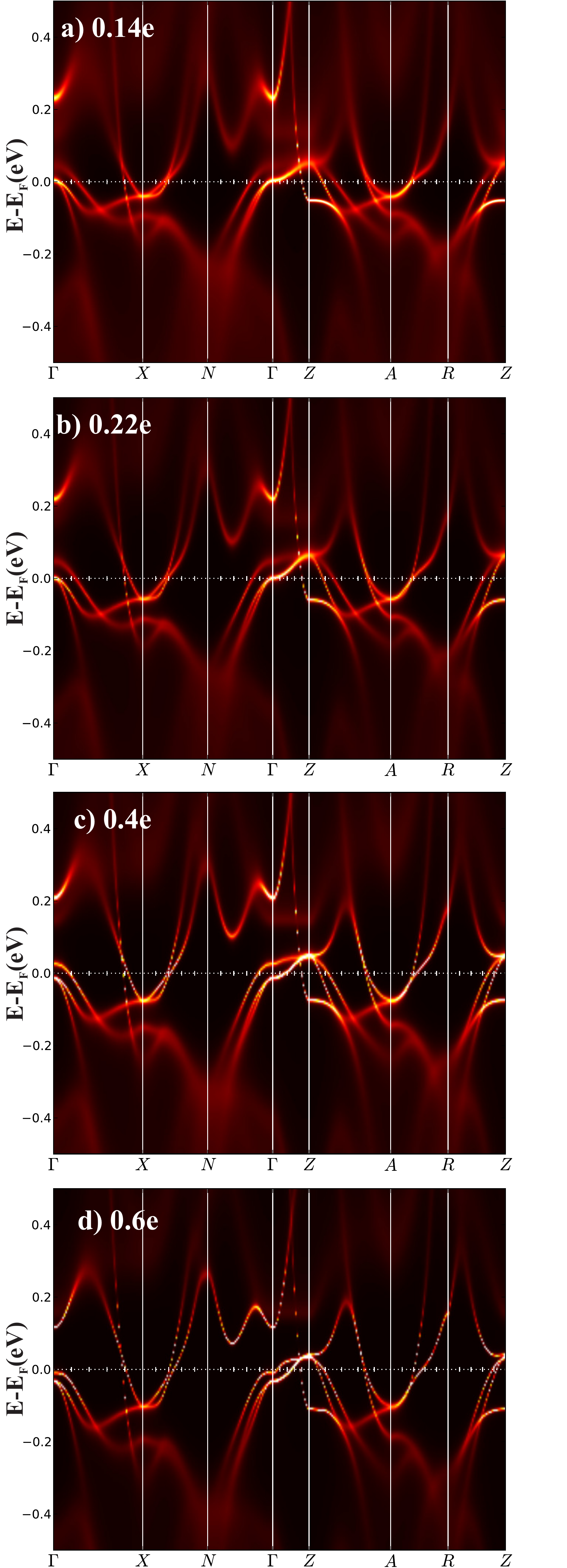}}
\caption{(Color online). The spectral functions of  BaFe$_{2-x}$Co$_x$As$_2$ computed by DFT+DMFT for various e-doping situations: (a) 0.14, (b) 0.22, (c) 0.4, and (d) 0.6 e/cell. The Brillouin zone with highsymmetry points is shown in Figure 1(f).
}

\end{figure}

At the zone corner, two elliptical electron pockets create the Fermi surface. The outer band has mostly the xy character and the inner band has mostly the xz/yz character. Adding holes to the system causes that the electron pockets lose some weight and become smaller. In the critical point, the inner band would not cross E$_F$ and a Lifshitz transition would happen in the system. This Lifshitz transition changes the Fermi surface at the zone corner to a tetramerous hole pocket, and now there is no electron pocket at the Fermi surface (‎Figure 7). As shown in ‎Figure 4, we obtained the critical point of $x=0.7$ for Ba$_{1-x}$K$_x$Fe$_2$As$_2$ in our calculations. This is consistent with the reported ARPES measurements [8,48]. Using DFT calculations, Khan and Jahnson have reported $x=0.9$ as the critical point that is not compatible with the existing experimental results. An abrupt change in the energy gap of the superconducting Ba$_{1-x}$K$_x$Fe$_2$As$_2$ 2 for $x\geq0.6$ and a jump in its heat capacity at Tc for  $x\approx0.7$ have been reported by Malaeb et al. [8] and Bud’ko et al. [7], respectively. Our findings suggests that the Lifshitz transition has caused the phenomena. However, the results show the major role of the electronic structure in these materials.

Okazaki et al.  [14] have reported for KFe$_2$As$_2$ an almost zero gap on the outer Fermi surface at the zone center, but superconductivity gaps on the middle and inner Fermi surfaces. It is unknown that why the outer band shows no superconductivity. To explore this more precisely, the spectral functions corresponding to the three situations of 0.8, 0.9, and 1 h/cell were plotted along the $\Gamma-X-Z$ path in ‎Figure 5. Our findings show the K$_z$ dependency in a band that its character mostly results from the z2 orbital (see ‎Figure 5, the blue band at  $\Gamma$). The band exists for 0.9 h, but it disappears for 1 h. However, for KFe$_2$As$_2$ , there is only two hole pockets at the $Z$ point, while three bands exist at the  $\Gamma$ point. The finding may explain the reason behind the almost zero gap on the outer Fermi surface at the zone center. The result emphasizes the major role of the electronic structure in the superconductivity of high-temperature superconductors. 

The spectral functions of BaFe$_{2-x}$Co$_x$As$_2$ computed by DFT+DMFT for various $x$ are shown in ‎Figure 6. Our calculations demonstrate at least two Lifshitz transitions when the e-doping amount increases, until $x=0.6$. A Lifshitz transition occurred between $ x=0.22$ and $x=0.4$, where the superconductivity state disappears [11]. Liu et al. [49] have reported a change in the thermoelectric power around $x=0.4$, which could be explained by the phenomenon. For $x=0.6$, there is no hole pocket around the zone center. The only hole pocket at this point for $x=0.4$ is removed by further e-doping, which creates the second Lifshitz transition for when the e-doping amount increases. Therefore, for $x\geq0.6$, only electron pockets exist $0.4< x<0.6$. Furthermore, the electron pocket at the zone corner expands and becomes very wide when e-doping increases. It seems that in the e-doping regime, E$_F$ increases smoothly by increasing the number of electrons in the system. The calculated spectral function has a good agreement with the reported ARPES measurement [49]. Our results show the importance of the Fermi surface topology for superconductivity in  BaFe$_{2-x}$Co$_x$As$_2$. Nevertheless, another tuning parameter, i.e. the structural parameter, may play an important role for superconductivity in the materials. Recently, Kyung et al. [10] have enhanced the superconductivity  in the surface-electron-doped iron-pnictide Ba(Fe$_{1.94}$Co$_{0.06}$)$_2$As$_2$ compound, with the same Fermi surface topology as here for $x=0.4$ , for which no superconductivity state exists in BaFe$_{2-x}$Co$_x$As$_2$. They have pointed out that the structural parameter must not be ignored. However, our calculations show the strength of the DFT+DMFT method in predicting the electronic structure of iron-based superconductors.

Schematics of the variation of the BaFe$_2$As$_2$ Fermi surface through electron-doping and hole-doping are illustrated in ‎Figure 7, where the upper panel shows the process of the Fermi surface variation as the h-doping increasing, derived from ‎Figure 3. The colors denote the orbitals that mostly characterize the pockets. Increasing hole-doping to the system expands the hole pockets and shrinks the electron pockets. The processes continues until the critical point, where electron pockets disappear completely and a tetramerous hole pocket appears at the zone corner. Our spectral function for Ba$_{1-x}$K$_x$Fe$_2$As$_2$ demonstrates the critical point between $x =0.6$ and $0.7$. At the zone center, the hole pockets gain some weight and expand when the hole-doping to the system increases. For KFe$_2$As$_2$ , there are three expanded hole pockets at the center that has a good agreement with ARPES measurement [14].  The outer one has the xy character and the middle and the inner pockets have mostly the xz/yz character. The lower panel of ‎Figure 7 is related to the e-doping regime. In contrast to the hole-doping, increasing the electron-doping expands the electron pockets and disappears the hole pockets from the Fermi surface. The results are consistent with the calculated ON, which indicates that the xz, yz and xy orbitals gain some carrier concentrations. E-doping enhances E$_F$ in the system and dissolves hole pockets at the Fermi surface. Our findings have a good agreement with the reported ARPES measurement [49].

\begin{figure}
\centerline{\includegraphics[width=1.0\linewidth]{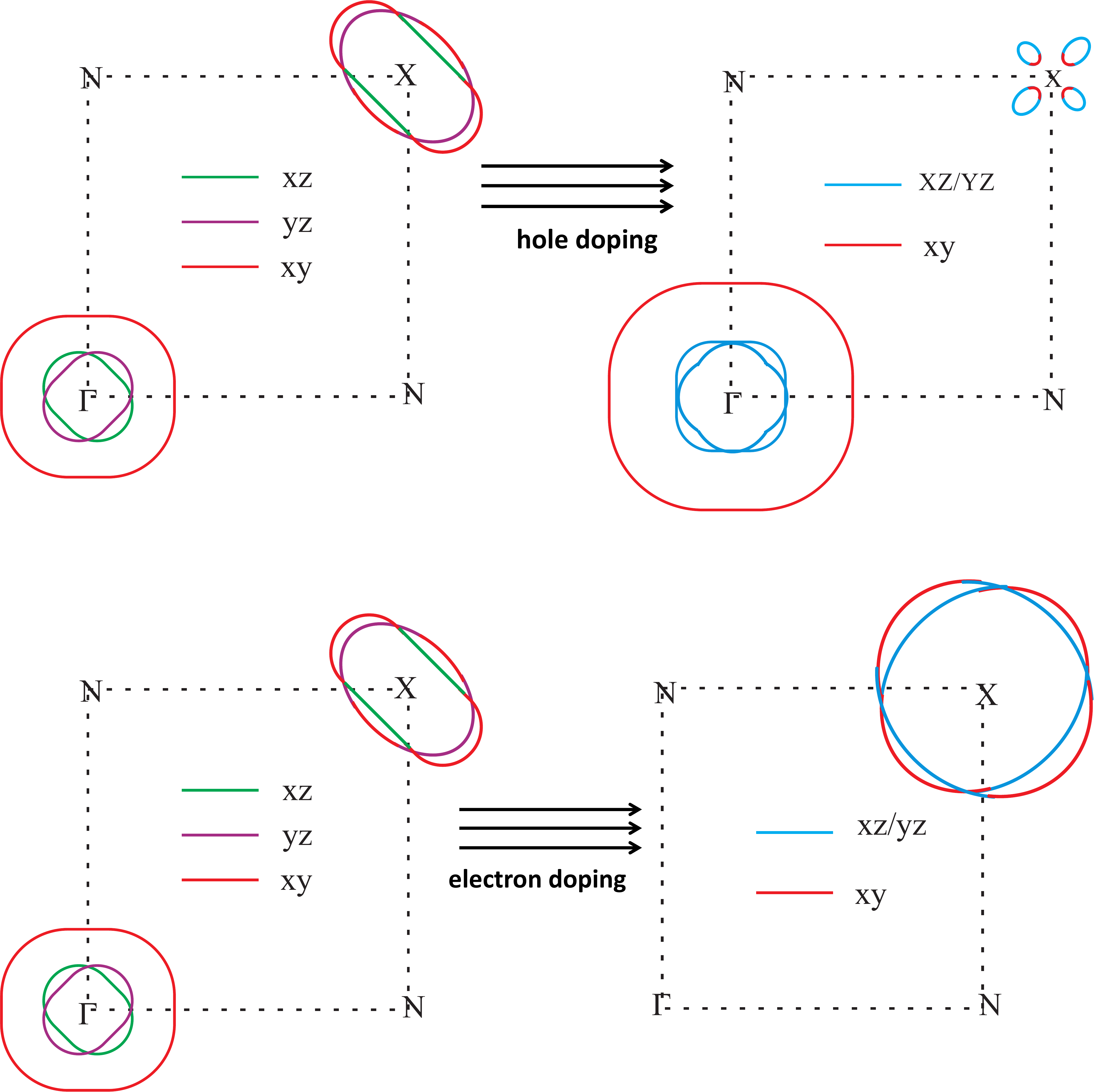}}
\caption{(Color online) Schematics of the variations of the BaFe$_2$As$_2$ Fermi surface through electron- and hole-doping. 
}

\end{figure}

The correlation strength in iron-based superconductors has always been under debate. Depending on their structural parameter, the materials are classified as weakly- to strongly-correlated. Pnictides are classified as weakly- or moderately-correlated, while chalcogenides are classified as moderately- or strongly-correlated [22]. Structural parameters, as the tuning parameters, especially the anionic height, play an important role in these classifications [50]. Recently, some theoretical and experimental evidence has predicted the strong correlation in KFe$_2$As$_2$ [21,23,46]. Here, we used the DFT+DMFT method as a realistic method to investigate BaFe$_2$As$_2$ for a wide range of electron- and hole-doping amounts. We found that in the heavily h-doped region, the xy orbital tends to its half-filled situations and become much more correlated than the other orbitals. This is because the orbital has a higher energy than the others and it is narrower near E$_F$. The properties make the xy orbital a good candidate for these behaviors. The evidence shows the tendency of the xy orbital to the orbital-selective Mott insulator behavior. There are rare systematic and realistic studies that could predict the evidence of the tendency of pnictides to OSMP. Our results show the strength of the DFT+DMFT method in predicting the correlation of the materials. 

The electronic structure is very important in the superconductivity features of high-temperature superconductors. Recently, the precise experimental technique of ARPES has made a great progress in determining the electronic structure and its effects on the materials. Finding a theoretical method that could predict the same ARPES results is very important, which enhances our understanding of the pairing mechanism. The spectral functions calculated by us through the DFT+DMFT method are consistent with the reported ARPES results [8,14,49], with a good precision for a wide range of electron- and hole-doping amounts.

\section{Conclusion}
In summary, we calculated the electronic structure of the electron- and hole-doped  BaFe$_2$As$_2$ compounds using the combined DFT+DMFT method. Our results showed the coherent-incoherent crossover when the h-doping varies around its optimum amount. The xy orbital was found much correlated than the other orbitals, which is because the orbital loses its carrier concentration and reaches its half-filled situation. This orbital becomes narrower around E$_F$ in the heavily h-doped region, which tends toward strongly-correlated systems. The phenomena suggest the tendency of the system toward the orbital-selective Mott phase. According to our findings, we can consistently believe that the strongly-correlated behavior in pnictides is due to their tendency towards OSMP. The results clarified the importance of the electronic correlation in these materials and the role played by the xy orbital in this matter. Our calculated spectral functions demonstrated that several Lifshitz transitions occur in electron- and hole-doped regions that could explain some critical behaviors across the regions: the abrupt change in the energy gap of superconducting  Ba$_{1-x}$K$_x$Fe$_2$As$_2$ at $x\geq0.6$, the jump in its heat capacity at T$_c$ for $x\approx0.7$, and the variation of the thermoelectric power of BaFe$_{2-x}$Co$_x$As$_2$ around $x=0.4$. The findings highlight the role of the electronic structure in the physical characterization of these materials. The spectral functions calculated by DFT+DMFT have a very good agreement with the reported ARPES measurements in the literature, and it emphasizes the strength of this method in predicting the properties of iron-based superconductors.

\section*{Acknowledgement}
We are very grateful to the Professor Kristjan Haule for providing the DFT+DMFT code and also his tutorials and responses to our questions and comments. We also would thank Dr. Gheorghe L. Pascut for his tutorials and comments. The calculation was performed in the high-performance computing (HPC) Center of Tarbiat Modares University (TMU).

\end{document}